# The generation of ground state structures and electronic properties of ternary $Al_kTi_lNi_m$ clusters ($k + l + m = 4$) from a two-stage DFT global searching approach

Pin Wai Koh[1], Tiem Leong Yoon[1,*], Thong Leng Lim[2,*], Yee Hui Robin Chang[3]

[1]School of Physics, Universiti Sains Malaysia, 11800 USM, Penang, Malaysia

[2]Faculty of Engineering and Technology, Multimedia University, Melaka, Malaysia

[3]Applied Sciences Faculty, Universiti Teknologi MARA, 94300 Kota Samarahan, Sarawak, Malaysia

[*]Corresponding authors

E-mail addresses: kohpinwai@gmail.com (P.W. Koh), tlyoon@usm.my (T. L. Yoon), tllim@mmu.edu.my (T. L. Lim), robincyh@sarawak.uitm.edu.my (Y. H. R. Chang)

## Abstract

Structural and electronic properties of ternary clusters $Al_kTi_lNi_m$, where $k$, $l$, and $m$ are integers and $k + l + m = 4$ are investigated. These clusters are generated and studied by performing a two-stage density functional theory (DFT) calculations using the SWVN and B3LYP functional exchange correlations. In the first stage, an unbiased global search algorithm coupled with a DFT code with a light exchange-correlation and smaller basis sets are used to generate the lowest energy cluster structures. It is then followed by further optimization using another round of DFT calculation with heavy exchanged correlations and large basis set. Electronic properties of the structures obtained via the two-stage procedure are then studied via DFT calculations. The results are illustrated in the form of ternary diagram. Our DFT calculations find that the stability of the cluster increases with the increase in the number of nickel atoms inside the clusters. Our findings provide new insight into the ternary metallic cluster through the structure, stability, chemical order and electronic properties studies.

## 1. Introduction

Nanoclusters, or atomic clusters, are normally defined as aggregates of atoms containing from a few to a couple of thousands of atoms. Compared to its bulk, electronic and structural behaviors of a cluster usually become unscalable when its size reaches the nanometer scale, due to the low dimensional and quantum confinement effects [1]. In addition, structural and electronic properties of small size clusters in many cases could strongly dependent on the number of atoms. There is no general rules or consensual theoretical models that can predict the geometrical structures or electronic properties of a generic cluster, even more so for small ternary metallic clusters. Therefore, their properties are hardly correlated with the number of atoms, although the properties of the large size clusters (which consist of few hundred to a few thousand atoms) are similar to the bulk material. [2–3]

Transition metal clusters have been intensively studied, both experimentally [4–9] and computationally [10–13, 14–33] due to their broad applications in catalysis [34–36], magnetic-recording materials [37] and biological applications, to name a few. For example, Fe-Co-Ni [38] and FeAlAu$_n$ ($n = 1 – 6$) [39] trimetallic clusters have been studied for their magnetic, electronic, and structural properties. In catalytic research, many studies show that various type of catalytic reaction [40] exist in binary clusters and nanoparticles, and enhancement of the catalytic activity can be achieved by adding a second chemical element into the existing single-element clusters.

There is also the proof that the catalytic activity of some ternary noble or metallic clusters is far more superior than the unary and binary clusters [41]. For trimetallic cluster studies, structural change has been observed for Au-Pd-Pt [42], Au-Cu-Pt [43] and Rh-Pd-Pd [44] trimetallic nanoparticles when investigating their structural stability via improved evolutionary algorithm. It is found that little fractions on the surface for Au-Pd-Pt and Au-Cu-Pt nanoparticles are distributed by the Pt atoms, even when its compositions fall below 10%. Au and Ag atoms exhibit stronger surface aggregation than Cu atom, whereas Cu and $X$ ($X$ = Au, Ag, Pd) atoms tend to segregate towards the surface. These simulations are in good agreement with experimental results [45].

Computations based on classical and semi-classical methods such as that use Gupta potential, Sutton-Chen potential and others indicate that the ground state structure of the small clusters prefers icosahedron while for large clusters, truncated octahedron and a truncated decahedral structure are favored [4]. Classical and semi-classical approaches can tackle and explain the structural evolution of a cluster but may fail if electronic effects from valence electrons of the atoms are taken into consideration [1, 46]. It is expected that the search for ground state configurations of transition metal clusters, classical and semiclassical approaches will lead to unreliable results, due to the presence of localized d orbitals [47–50]. Simulation results fluctuate with different Density Functional Theory (DFT) software and optimization method used [47–50]. On the other hand, electronic structure and stability in transition metal clusters, especially 13-atom clusters of 3d/4d series, have been studied extensively by DFT methods in the last two decades [19–33]. Results from different DFT calculations differ due to the various type of exchange-correlation (XC) functional and basis set employed in the calculation, and the approach applied to sample candidate structures from the potential energy surface (PES). Different structures for $Ag_{13}$ and $Cu_{13}$ have also been reported from studies using either Gaussian orbital or plane-wave based DFT [27–28, 31]. DFT results also differ with inclusion or non-inclusion of the semi-core state in the pseudopotential [37, 51].

Study on the structure and bonding of aluminum clusters [52] has started since the 90s. Several experimental and theoretical works on magnetic properties [53], ionization and reactivities [54] and static polarizabilities [55] on aluminum clusters have been performed by Kaldor et al. in the 80s. A transition of non-jellium to jellium in aluminum clusters occurred when the cluster size exceeds 40 atoms [55]. Strong magnetic behavior is observed in dimers and some of the small aluminum clusters such as $Al_3$, $Al_6$, $Al_7$ and $Al_8$ [53]. Since the turn of the millennium, experimental and computational methods [56–60] are widely used to study the geometric, electronic and magnetic properties of the titanium clusters. As a typical transition element, titanium atom has a large number of vacant valence d orbitals, giving rise to a highly delocalized characteristic. Titanium clusters show oscillatory magnetism when the cluster size is small, while it is a nonmagnetic material in bulk [61]. Among all the 3d transition metal clusters, nickel clusters are widely studied experimentally and theoretically [51, 62 – 68]. The physical, chemical, electronic and magnetic properties of nickel clusters are closely related to their geometries, i.e. properties of Ni cluster changed with the number of nickel atom, and geometry also made an impact to the transition from diatomic to the bulk [69]. There are several theoretical and experimental studies on the Al-Ti, Al-Ni and Ni-Ti binary alloy systems (i.e. Al-Ti [70, 71], Al-Ni [72, 73] and Ni-Ti [74, 75] alloy systems) and ternary alloy system Al-Ti-Ni [76 – 78]. These binary and ternary alloy clusters are believed to be the potential catalyst of industrial engineering revolution [93]. So far, a few theoretical works on the small size binary alloy clusters involving Al, Ti and Ni atoms have been identified. Hua et al. [47] studied, with DFT, titanium-doped aluminum cluster Al$_n$Ti from 2 to 24 aluminum atoms. Also, within DFT framework, small bimetallic Ti-Ni clusters with total atom less than 13 are studied by Chen et al. [48], and Zhao et al. [49] work on the Al-Ni cluster with total atom less than 5 by using DFT method. On the other hand, the literature on the global search and generation of ground state structures of trimetallic clusters that are fully quantum-mechanically is very scarce. Structural and electronic properties of Al$_k$Ti$_l$Ni$_m$ ($k + l + m = 2, 3, 4$) clusters have been investigated by Erkov and Oymak [50, 79]. These authors generate the cluster structures based on a molecular dynamics (MD) scheme that uses Lennard-Jones (for two body part) and Axiltod-Teller triple-dipole potentials (for the three-body part) [80]. The electronic properties of the obtained structures are evaluated via DFT calculations within the Becke three-parameter, Lee-Yang-Parr (B3LYP) and effective core potential level. The search for the ground state structure of the Al-Ti-Ni clusters by Erkov et al. [50] is performed at the empirical level instead of the DFT level.

In this paper, our objectives are to generate, and then investigate, the ground state structures of Al$_k$Ti$_l$Ni$_m$ clusters for $k + l + m = 4$, using first-principle method. To obtain the ground state structures, a two-stage computational strategy will be deployed, where both stages will involve density functional theory calculations subjected to exchange-correlation functionals of different computational cost. A global minimum search algorithm, namely, basin-hopping (BH) will also be incorporated as an integral part of the two-stage computational strategy. Results from the first-principles calculations of geometric, chemical order and electronic properties will also be presented. Variation of the calculated properties with different stoichiometry is displayed in ternary diagrams.

The two-stage algorithm proposed in this paper has a practical advantage over other unbiassed global minimum search for the ground state structures of clusters, especially for multi-element ones. Quite generally, an unbiased search algorithm for the ground state structures of clusters involve the use of 'energy calculator' for calculating the total energy of a configuration during the search process. In the two-stage algorithm proposed in the paper, DFT is used as the energy calculator. In other more common used unbiased search algorithms, molecular dynamics or density-functional tight-binding (DFTB) are used as the energy calculator (at least partly). Search algorithms using DFTB or MD energy calculators are faster than that using only DFT (which is very uncommon). However, MD or DFTB energy calculators require the availability of empirical potentials (for MD) or Slater-Koster files (for DFTB). If the particular empirical input for a multi-element cluster is absent, such type of search algorithms become unfunctional. Employing only DFT as the sole energy calculator, which is first-principles in nature, the two-stage algorithm proposed in this work will not suffer drawback.

The article is organized as follows: Section 2 briefly describes the computational methods used in this work. In Section 3, low-energy structures and geometries are presented and energetics, stabilities and electronic properties of the trimetallic $Al_kTi_lNi_m$ clusters are discussed. The conclusion is drawn in Section 4.

## 2. Computational Methods

Basin-hopping is an unbiased optimization approach introduced by D. J. Wales and Doye [11–12] and has been widely employed in numerous theoretical works to locate the ground state structure or a global minimum of an atomic cluster system. The main idea of the BH approach is to transform a given PES into multidimensional staircase topography without altering the relative energies of global minimum and local minima. The transformed PES is given by

$$\tilde{V}(X) = \min\{V(X)\} \tag{1}$$

where the local energy minimization of a certain point $V(X)$ in PES is represented by min and $X$ is a set of coordinates $\{\boldsymbol{r}_1, \boldsymbol{r}_2 \ldots \ldots \boldsymbol{r}_N\}$ of $N$ atoms.

The initial configuration in a global search algorithm will strongly affect the ability to find the global minimum in the complex PES. There is a high possibility for an initial configuration to be trapped in a local minimum with a high energy barrier. Hence, it is advisable that the search algorithm is initiated with a series of initial configurations. In genetic algorithm (GA) method, optimization starts with the initialization of a population of initial guesses (individuals) which also known as "parents". All these individuals are spread randomly in a search space. A selection process is performed by determining the fitness on each of these individuals, and therefore those individuals with poor quality will be discarded and the remaining individuals will be kept for the next generation. Each "parents" that retained from selection process are subjected to genetic operations. The "child-breeding" and selection processes are repeated until the best individuals are obtained and the global minimum is supposedly contained in this collection.

The combination of basin-hopping and genetic algorithm methods, a novel search algorithm first introduced by Hsu and Lai [81], are employed as an optimization approach in this work. In this paper, a self-developed code by Lai and Hsu, known as parallel tempering multicanonical basin-hopping plus genetic algorithm (PTMBHGA), is deployed to generate the configurations of Al-Ti-Ni trimetallic clusters. In this work, PTMBHGA is interfaced with the first-principles code Gaussian 09 (G09) [82] for structural optimization and is dubbed as PTMBHGA-G09 hereafter. Details of the workings of PTMBHGA can be found in Refs. [81 - 84]. Here we outlined those modifications we introduced into the original PTMBHGA code.

In order to efficiently search for lowest energy structures of Al-Ti-Ni clusters using only DFT as the energy calculator throughout, a two-stage procedural strategy is introduced. Initially, two optimization algorithms, basin hopping (BH) and genetic algorithm (GA) are applied to generate the low-lying structures (LLS) within the density functional theory (DFT) potential energy surface (PES). 20 initial configurations are first randomly generated. BH is performed on each of these configurations (also known as candidates) for 100 steps. In each BH step, either the *angular move* or *random displacement* (AMRD) genetic-like operation [11–12, 81, 84–85] is applied. ARMD is a random move method to adjust the positions of the cluster structure and thus give birth to a new configuration. Another optimization operator called *cut and splice* genetic operation (GO), is also introduced. Cut and splice is a technique that employs a mating operator as a genetic operator to generate new structure configurations from a previous structure. The position of new cluster structure that generated by using either ARMD or cut and splice GO technique in each BH step is relaxed and its energy is calculated by using the G09 code. BH makes use of the limited-memory Broyden-Fletcher-Goldfarb-Shanno algorithm (L-BFGS) as the local energy minimization algorithm to minimize the potential energy of cluster structures. DFT calculations for the Al-Ti-Ni clusters are carried out using the Slater, Vosko, Wilks, and Nusair (SVWN) exchange-correlation functional and 3-21G Pople basis set, where SVWN is an exchange-correlation functional equivalent to local spin density approximation (LSDA) in G09. Next, these final 20 candidates from the previous 100 BH steps would be used as next generation "parents" to breed the "offspring" by using the cut and splice operator. *In the next generations*, five candidates with the lowest fitness value [71, 75] are removed and the remaining of the 15 candidates will once again be subjected to ARMD or cut and splice GO to generate another five new "offspring", replacing those candidates already been discarded, and thus the number of candidates in population size is maintained as 20. An "offspring" is optimized with 100 BH steps for each generation and is continued for 5 generations. Finally, 100 BH steps are performed on these 20 candidates independently again to guarantee the cluster structure with the lowest energy value is obtained.

At the second stage, the cluster's configuration with lowest energy from the first stage is once again subjected to the PTMBHGA-G09. Then, Monte Carlo BH is applied to this individual for 100 BH steps. Similar to the procedure of the first stage, this individual is subjected either to the techniques from Monte Carlo BH or GA to form new configurations. Differences between the first and second stage are that in the first stage, clusters are randomly generated, a low-quality XC correlation and basis set in DFT are used to explore the PES of the clusters.-For the second stage, candidates with the lowest energy from the first stage would undergo the same procedures as in the first stage but by using a higher quality XC correlation and basis set in the DFT calculations. These 100 cycles of geometry optimization follow the Berny optimization procedures [86] and in self-consistent field (SCF) calculation, Becke three-parameter, Lee-Yang-Parr (B3LYP) exchange-correlation functional with 6–311G* basis set in G09 is employed. Vibrational frequency calculations with the ultrafine grid are also carried out for each cluster. Frequency calculations are conducted to ensure optimized structures are located at the minima instead of transition states.

## 3. Results and Discussion

### 3.1 Structure

It is found that one-third of the cluster structures investigated are 2D planar, i.e. square or parallelogram, while the rest of the cluster is equipped with trigonal pyramid-like structure except the $Ti_1Ni_3$ cluster, which is highly nonsymmetrical. For pure clusters case, $Al_4$ has a parallelogram-like geometry that agrees well with those results published by Jones [52] while $Ti_4$ cluster possesses tetrahedron shape, which agrees with Sun et al. [61]. Lowest total energy $Ni_4$ cluster with -164160.048 eV, acquired by using the two-stage DFT method has a square-like geometry, comparable with the work from Khanna et al. [68]. Isomer for $Ni_4$ cluster, claimed by Erkov et al. [50] and Goel et al. [69] has a parallelogram-like structure, and we have found the same structure with an energy of -164159.170 eV. For bimetallic clusters, our simulation result shows the same $Al_3Ti_1$ cluster as investigated by Hua et al. [47] which employs the tight binding genetic algorithm (TBGA) combined with DFT. For Ti-Ni system, our work delivers the same result as those work done by Chen et al. [48], except for the $Ti_1Ni_3$ cluster. This structure, appearing in the shape of a Y character (see ref. Fig. 3 in [48]), is found to be a ground state structure in this work, whereas it appears as an isomer in ref. [48]. However, our work on Al-Ni structure has NO similar structures as compared to that reported by Zhao et al. [49], where the authors have used a crude DFT method to generate ground state structures for the Al-Ni clusters. When our results on $Al_kTi_lNi_m$ clusters are compared with those investigated by Erkov et al. [50], which uses self-parameterize many-body empirical potential implemented in a molecular dynamics (MD) approach, most optimized structures that we have found are totally different. Generally, for small binary and ternary clusters consist of the combination of Al, Ti, Ni atoms, there are no experimental data to compare with. In Fig. 1, ground state structures of the four-atom Al-Ti-Ni clusters are illustrated. In Fig. 2, average interatomic distance (AID) of the cluster as a function of the atomic composition is shown and the AID values for each cluster are collected and tabled as in Table 1. Average interatomic distance for $Ni_4$ is smaller than $Ti_4$ whereas $Al_4$ has the largest AID among all these pure clusters and also the second largest AID candidate in the $Al_kTi_lNi_m$ system. Comparing among binary clusters, Al-Ti cluster has the largest AID especially $Al_3Ti_1$ which has a slightly larger interatomic distance than $Al_4$. In the Al-Ni and Ti-Ni binary clusters, the AID of these clusters decrease as the number of nickel atom inside the system increases. For instance, the AID value of the $Ti_1Ni_3$ is lower than the homonuclear cluster $Ni_4$, which is the cluster with second shortest interatomic distance in the system. For Al-Ti-Ni ternary case, an AID of the $Al_2Ti_1Ni_1$ is shorter than $Al_1Ti_2Ni_1$ and $Al_1Ti_1Ni_2$ has the shortest AID among all the Al-Ti-Ni ternary clusters.

### 3.2 Stability

The binding energy per atom $(E_b)$ of a cluster like $Al_kTi_lNi_m$ with size $k + l + m = 4$ is calculated by using the equation as follows:

$$E_b(Al_kTi_lNi_m) = \frac{E(Al_kTi_lNi_m) - kE(Al) - lE(Ti) - mE(Ni)}{k + l + m}. \qquad (2)$$

In general, the total thermodynamic stability of a specified cluster can be measured through its binding energy per atom. A cluster [1, 40] would be considered more stable when its binding energy per atom is more negative. In Fig.3, $E_b$ as the function of whole cluster composition is presented in a simple ternary diagram. It is noticed that binding energies are larger for the binary Al-Ni and Ti-Ni regions, especially when the Al composition is reduced, or the Ni composition is increased. Pure $Al_4$ cluster retains the smallest binding energy. The number of Ni atoms is the main factor that dominates the binding energy of $Al_kTi_lNi_m$ cluster. Binary and ternary clusters with higher Ni concentration such as $Al_1Ni_3$, $Ti_2Ni_2$ and $Al_1Ti_1Ni_2$ clusters would display larger $E_b$, i.e., 2.84, 2.81 and 2.83 eV respectively. Largely speaking, cluster with higher stability exhibit higher value of $E_b$.

Calculations of the excess energy $(E_{exc})$ and the second difference energy $\Delta_E$ for the ternary clusters are carried out to further determine the cluster formation with a given composition, which is also known as possible magic compositions. Ferrando et al. [40] have worked on these quantities for the case of binary clusters and later they are generalized by Granja et al. for the ternary cluster [1, 40]. $E_{\text{exc}}$ and $\Delta_E$ can be calculated for the system $\text{Al}_k\text{Ti}_l\text{Ni}_m$ $(k + l + m = N)$ as follows:

$$E_{\text{exc}}(\text{Al}_k\text{Ti}_l\text{Ni}_m) = E_b(\text{Al}_k\text{Ti}_l\text{Ni}_m) - k\frac{E_b(\text{Al}_N)}{N} - l\frac{E_b(\text{Ti}_N)}{N} - m\frac{E_b(\text{Ni}_N)}{N}, \qquad (3)$$

$$\begin{aligned}
\Delta_E(\text{Al}_k\text{Ti}_l\text{Ni}_m) = \frac{1}{n_{klm}} &E_b(\text{Al}_{k+1}\text{Ti}_{l-1}\text{Ni}_m) \\
&+ E_b(\text{Al}_{k-1}\text{Ti}_{l+1}\text{Ni}_m) + E_b(\text{Al}_{k+1}\text{Ti}_l\text{Ni}_{m-1}) \\
&+ E_b(\text{Al}_{k-1}\text{Ti}_l\text{Ni}_{m+1}) + E_b(\text{Al}_k\text{Ti}_{l+1}\text{Ni}_{m-1}) \\
&+ E_b(\text{Al}_k\text{Ti}_{l-1}\text{Ni}_{m+1}) - n_{klm}E_b(\text{Al}_k\text{Ti}_l\text{Ni}_m)
\end{aligned} \qquad (4)$$

where $E_b$ is denoted as the binding energy of the corresponding cluster, while $n_{klm}$ and $1/n_{klm}$ is the total number of the nearest neighbor structure and normalization factor of a cluster, respectively. To ensure a better comparison of $\Delta_E$ values for pure, binary and ternary clusters, the normalization factor of two is used for the pure clusters, increases to four for the binary ones and to six for the ternary clusters.

Note that pure clusters have zero excess energy, $E_{\text{exc}} = 0$ and cluster with a negative $E_{\text{exc}}$ value implies that mixing is preferable, i.e. ternary and binary clusters are more preferable than pure $\text{Al}_N$, $\text{Ti}_N$, and $\text{Ni}_N$ in cluster formations. Maximum $\Delta_E$ value also indicates that the cluster possesses high relative stability; i.e. a magic composition. In Figs. 4 and 5, values obtained for $E_{\text{exc}}$ and $\Delta_E$ are plotted to display the stability of four-atom clusters (refers to Fig. 1). As the number of heteronuclear bonds of the cluster increases, its stability also increases, which reflects elements in the cluster tends to be mixed rather than segregated. From the Figs. 4 and 5, both $E_{\text{exc}}$ and $\Delta_E$ values exhibit similar trend with the composition. It is found that the $E_{\text{exc}}$ and $\Delta_E$ tend to display larger values for the binary Al-Ni and Ti-Ni and the ternary compositions in this order. For example, in Co-Ni clusters, they are very near to the pure Ni and Co compositions in the periodic table and are the least preferred candidates for alloy characterization [38, 46 ,87–88]. Comparing the minima obtained for $E_{\text{exc}}$ and maxima for $\Delta_E$, the most favorable binary and ternary alloys are $Al_1Ni_3$ and $Al_1Ti_1Ni_2$. In fact, these alloys are found abundantly in the cluster growing experiment [38]. The values of binding energy per atom, excessive energy, and second order difference energy are tabled in Table1.

### 3.3 Chemical Order

In order to understand the mutual influence of the multi-component alloy structure, the chemical order (segregation or mixing) for all the $Al_kTi_lNi_m$ cluster configurations is studied. Ducastelle [88] introduced chemical order to study bulk like binary alloy system. Bulk-like binary alloy systems displays a clear distinction between disorder and mixing when its chemical order, $\sigma$, approximate zero and small negative, respectively. Ordered phases such as layered-like phase may appear in the bulk-like binary alloy systems when $\sigma$ is a large negative value [88]. Chemical order $\sigma$ as a function of the relative composition has the followings characteristics: positive value when segregation or phase separation takes place, which also means that cluster is favored to form homo–atomics pairs; negative when mixing occurs, which also indicates that hetero-atomic pairs are more prominent in the cluster. If the chemical order value decreases towards zero, the cluster is said to undergo a phase transition from segregation to mixing.

Based on several literature reviews [46, 87 − 88], the chemical order parameter ($\sigma$) in our case can be defined as follows:

$$\sigma = \frac{n_{\text{Al−Al}} + n_{\text{Ti−Ti}} + n_{\text{Ni−Ni}} − n_{\text{Al−Ti}} − n_{\text{Al−Ni}} − n_{\text{Ti−Ni}}}{n_{\text{Al−Al}} + n_{\text{Ti−Ti}} + n_{\text{Ni−Ni}} + n_{\text{Al−Ti}} + n_{\text{Al−Ni}} + n_{\text{Ti−Ni}}}, \tag{5}$$

where $n_{A-B}$ is the number of nearest $A-B$ bonds (see the pair distribution on Table 2). The $\sigma$ value for each cluster is displayed in Table 2.

In Fig. 6, the order parameters $\sigma$ for all the cluster configurations are given. As expected, segregation ($\sigma = 1$) is observed near to the corner of the triangle, i.e., in the region where all the pure elements clusters are located, e.g. $Al_4$, $Ti_4$, and $Ni_4$ clusters. Bimetallic clusters $Al_3Ti_1$, $Al_1Ti_3$, $Al_3Ni_1$ and $Ti_3Ni_1$ display zero order parameter, indicating a transition between segregation and mixing. The mixing phase (negative $\sigma$) located mainly at the central region of the triangle (also known as a ternary region), suggesting that ternary clusters are more favorable on mixing and they are more stable with more heterogeneous bonds.

### 3.4 Electronic Properties

Ionization potential (IP) and electron affinity (EA) for $Al_kTi_lNi_m$ clusters are illustrated in Figs. 7 and 8. IP is computed as the total energy difference between the electronic ground structure of the cationic and the neutral cluster structures, whereas the total energy difference between the neutral and anionic cluster structures is known as EA. Both IP and EA of the cluster are defined as

$$\text{IP} = E_{Al_kTi_lNi_m} - E_{Al_kTi_lNi_m^+},\tag{6}$$

$$\text{EA} = E_{Al_kTi_lNi_m^-} - E_{Al_kTi_lNi_m}.\tag{7}$$

A cluster with higher IP implies that it is more difficult to remove electrons from the neutral cluster. A cluster with higher EA implies that huge amount of energy is released when an electron is added to a neutral cluster. The values of IP and EA obtained are then used to calculate global hardness ($\eta$) and Mulliken electronegativity ($\xi$) for all the cluster configurations and they are shown in Fig. 9 and 10. Both parameters are crucial quantitative parameters that employed to measure the chemical reactivity of a given specific cluster in a charge transfer process and are defined as

$$\eta = \mathrm{IP} - \mathrm{EA}\,, \tag{8}$$

$$\xi = \frac{1}{2}(\mathrm{IP} + \mathrm{EA})\,. \tag{9}$$

High values of IP are observed in the domain close the line connecting Al and Ni (Fig. 7) and $Al_1Ni_3$ cluster has the highest IP among all the clusters. High IP also implies increased difficulty in removing an electron from the $Al_1Ni_3$ cluster to form a cationic cluster. However, IP of all the ternary clusters is slightly lower when compared to the pure element and binary clusters. The entire pure element clusters exhibit high EA values. High values of EA are also displayed along the Ti-Ni edge. $Al_1Ni_3$ cluster is the only cluster which exhibits a negative value of EA. This indicates that $Al_1Ni_3$ cluster is highly unstable to form an anionic cluster when an electron is added to it. $Al_1Ni_3$ cluster also possesses the highest value of global hardness in the system. The Mulliken electronegativity is correlated to the chemical potential ($\mu$) of the system. The largest electronegativities are in the vicinity of pure $Al_4$ and $Ni_4$ clusters. Electronic data that includes ionization potential, electron affinity, global hardness and Mulliken electronegativity for the Al-Ti-Ni cluster system are reported in Table 2.

Molecular orbital energy gap ($E_{\mathrm{gap}}$) is defined as an energy difference between the HOMO (highest occupied molecular orbital) and the LUMO (lowest unoccupied molecular orbital). The ability of an electron to transfer from an occupied orbital to an unoccupied orbital is measured by the HOMO-LUMO energy gaps. Referring to the work by Sansores et al. [89], the overall $E_{\mathrm{gap}}$ is defined by

$$E_{\sigma\sigma'} = |E(\text{HOMO}_{\sigma'}) - E(\text{LUMO}_{\sigma})| \sigma, \sigma' = \alpha, \beta, E_{\text{HOMO-LUMO}} = \min\{E_{\sigma\sigma'}\}. \quad (10)$$

For closed shell systems, $\text{HOMO}_\alpha$ and $\text{LUMO}_\alpha$ according to spin are calculated and the spin value for $\text{HOMO}_\alpha$, $\text{LUMO}_\alpha$, $\text{HOMO}_\beta$ and $\text{LUMO}_\beta$ are calculated for the opened shell systems. Cluster with a high value of $E_{\text{gap}}$ indicates that it is weak in chemical reactivity. In Fig. 11, $E_{\text{gap}}$ for all the $\text{Al}_k\text{Ti}_l\text{Ni}_m$ clusters are illustrated in the form of the ternary diagram. The bimetallic $\text{Al}_1\text{Ni}_3$, $\text{Al}_2\text{Ni}_2$, and $\text{Ti}_1\text{Ni}_3$ clusters exhibit larger $E_{\text{gap}}$ along the Al–Ni and Ti–Ni edges. The larger values of the $E_{\text{gap}}$ are also observed for all the ternary clusters, which reaffirm that the clusters are stable with respect to alloying. The value of $E_{\text{gap}}$ for each cluster can also be found in Table 1.

| Alloy | Symmetry | AID | $E_b$ | $E_{exc}$ | $\Delta_E$ | $E_{gap}$ |
|-------|----------|-----|-------|-----------|------------|-----------|
| 004 | $D_4h$ | 2.1470 | −2.0367 | 0 | −0.7405 | 1.8447 |
| 040 | $D_2d$ | 2.4147 | −2.3344 | 0 | −0.1313 | 1.4291 |
| 400 | $D_2d$ | 2.6378 | −1.1871 | 0 | −0.7757 | 1.1551 |
| 013 | $C_s$ | 2.0803 | −2.7141 | −0.6028 | 0.0839 | 2.1970 |
| 022 | $C_2v$ | 2.2196 | −2.8125 | −0.6269 | 0.1131 | 1.5703 |
| 031 | $C_s$ | 2.3652 | −2.5400 | −0.2798 | −0.0232 | 1.6819 |
| 103 | $C_s$ | 2.3264 | −2.8406 | −1.0161 | 0.3115 | 2.3818 |
| 202 | $C_2v$ | 2.3093 | −2.5348 | −0.9228 | −0.0197 | 2.1070 |
| 301 | $C_s$ | 2.4697 | −2.1889 | −0.7893 | 0.2346 | 1.7124 |
| 130 | $C_s$ | 2.6020 | −2.3916 | −0.3440 | −0.0280 | 1.8332 |
| 220 | $C_2v$ | 2.4895 | −2.0905 | −0.3297 | −0.2097 | 1.7636 |
| 310 | $C_s$ | 2.6940 | −1.7368 | −0.2628 | −0.2195 | 1.5211 |
| 112 | $C_2v$ | 2.4308 | −2.8303 | −0.9315 | 0.1679 | 2.0379 |
| 121 | $C_1$ | 2.4910 | −2.7137 | −0.7405 | 0.2098 | 2.0917 |
| 211 | $C_s$ | 2.2882 | −2.3586 | −0.6722 | 0.0094 | 1.9328 |

Table 1. Properties of $Al_kTi_lNi_m$ clusters with $k + l + m = 4$. Numbers in the first column indicate the number of Al, Ti and Ni atoms in each cluster. The average interatomic distances (in Å), binding energy per atom ($E_b$ in eV), excess energy ($E_{exc}$ in eV), second order difference energy ($\Delta_E$ in eV), and HOMO–LUMO energy gaps ($E_{gap}$ in eV) are presented in the table.

| Alloy | $N-N$ pairs | $\sigma$ | IP | EA | $\eta$ | $\xi$ |
|-------|-------------|----------|------|------|--------|-------|
| 004 | 0 0 4 0 0 0 | 1.00 | 6.3540 | 2.0559 | 4.2980 | 4.2049 |
| 040 | 0 6 0 0 0 0 | 1.00 | 4.7438 | 1.6283 | 3.1155 | 3.1861 |
| 400 | 5 0 0 0 0 0 | 1.00 | 6.4327 | 2.0634 | 4.3693 | 4.2480 |
| 013 | 0 0 1 0 0 3 | −0.50 | 5.9940 | 1.9852 | 4.0088 | 3.9896 |
| 022 | 0 1 0 0 0 4 | −0.60 | 4.7672 | 1.0979 | 3.6693 | 2.9326 |
| 031 | 0 3 0 0 0 3 | 0 | 4.9452 | 1.9006 | 3.0446 | 3.4229 |
| 103 | 0 0 2 3 0 0 | −0.20 | 8.0000 | −0.9729 | 8.9729 | 3.5136 |
| 202 | 1 0 1 4 0 0 | −0.33 | 6.3200 | 1.4594 | 4.8606 | 3.8897 |
| 301 | 3 0 0 3 0 0 | 0 | 6.0230 | 1.2477 | 4.7752 | 3.6353 |
| 130 | 0 3 0 0 3 0 | 0 | 6.2092 | 0.5180 | 5.6912 | 3.3636 |
| 220 | 1 1 0 0 4 0 | −0.33 | 5.8590 | 1.7686 | 4.0905 | 3.8139 |
| 310 | 3 0 0 0 3 0 | 0 | 6.5509 | 1.0697 | 5.4812 | 3.8103 |
| 112 | 0 0 0 2 1 2 | −1.00 | 5.6301 | 0.8387 | 4.7914 | 3.2344 |
| 121 | 0 1 0 1 2 2 | −0.67 | 6.6626 | 0.4201 | 6.2424 | 3.5414 |
| 211 | 1 0 0 2 2 1 | -0.67 | 5.3713 | 1.7628 | 3.6086 | 3.5671 |

Table2. Properties of $Al_k Ti_l Ni_m$ clusters with $k+l+m=4$. Numbers in the first column indicate the number of Al, Ti and Ni atoms in each cluster. The number of nearest neighbor pairs (the order is Al–Al, Ti–Ti, Ni–Ni, Al–Ni, Al–Ti and Ti–Ni, respectively), chemical order ($\sigma$), ionization potential (IP in eV), electron affinity (EA in eV), global hardness ($\eta$ in eV) and Mulliken electronegativity ($\xi$ in eV) are presented in the table.

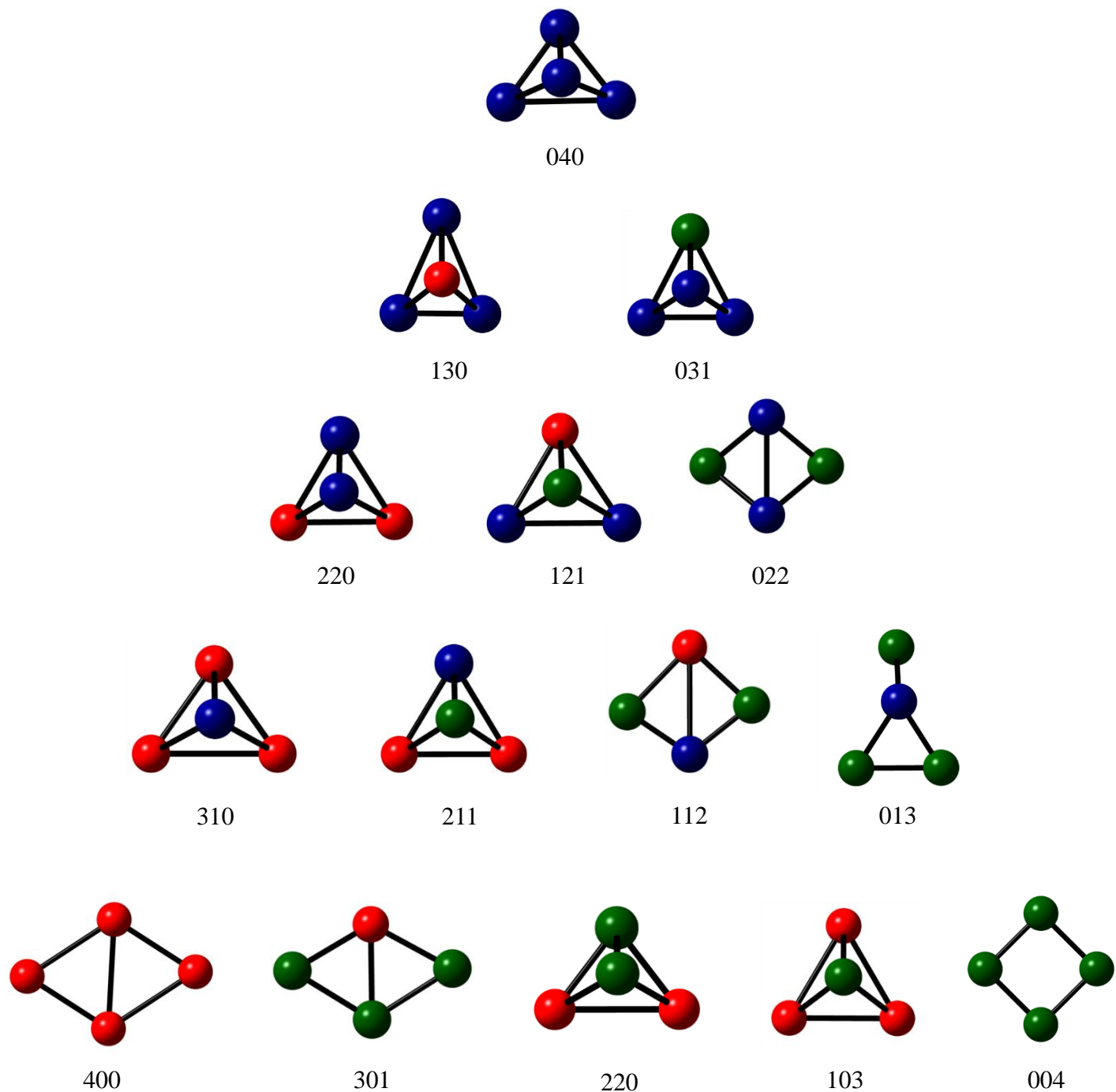

Fig. 1 Ground states structures of $Al_kTi_lNi_m$ ($k + l + m = 4$) clusters as a function of the atomic composition. Blue sphere represents Ti atoms, red for Al and green for Ni. Number below each model of cluster geometry indicates the number of Ti, Ni and Al atoms in each cluster.

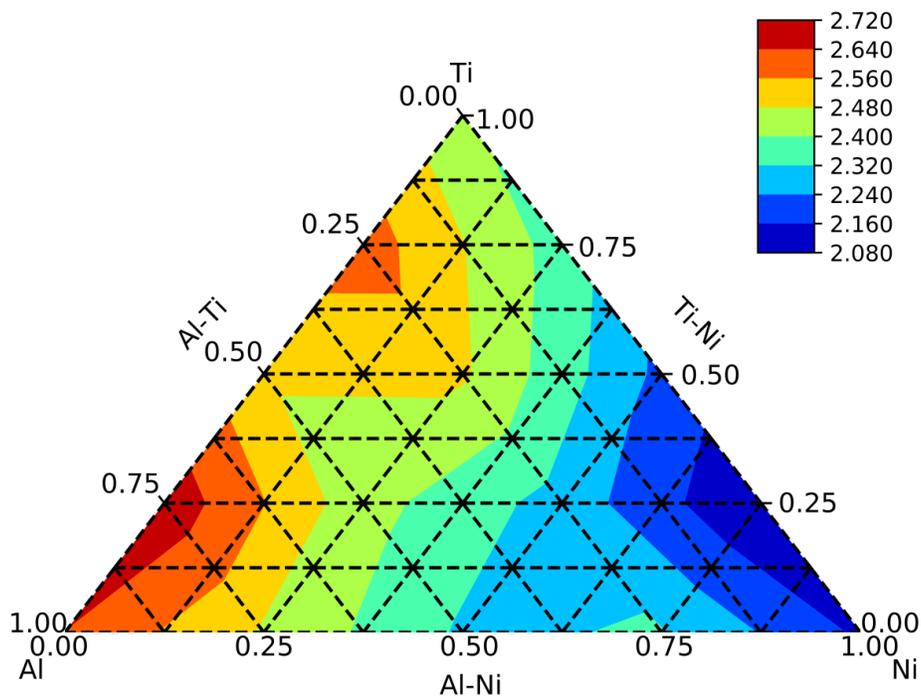

Fig. 2 Average interatomic distance (in Å) of $Al_kTi_lNi_m$ ($k + l + m = 4$) clusters as a function of the atomic composition.

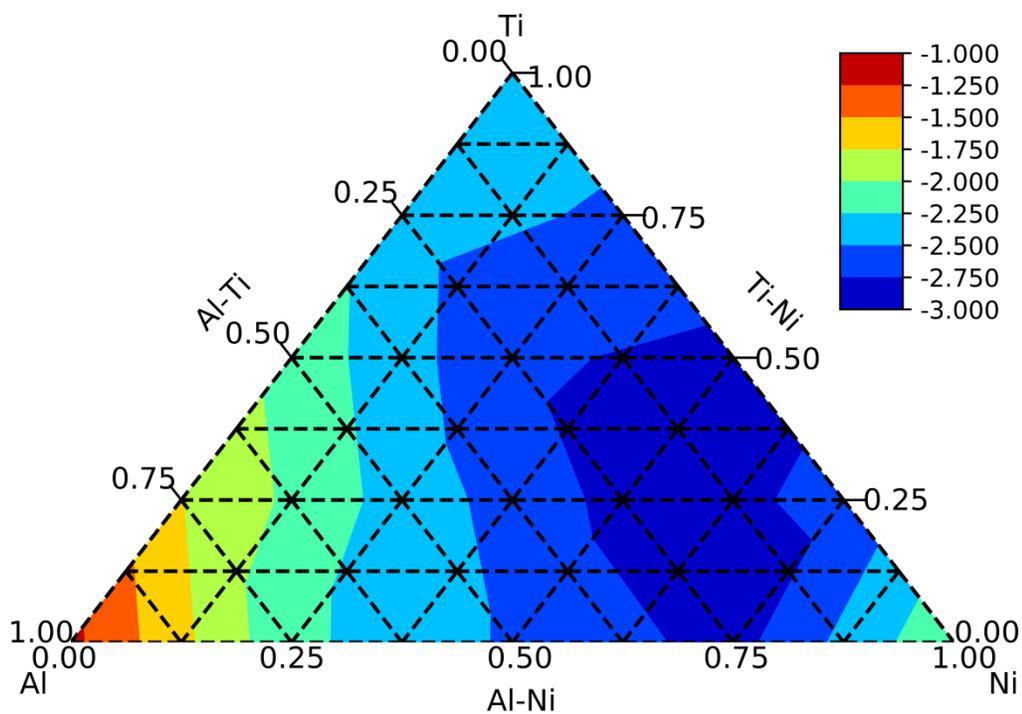

Fig. 3 Binding energy per atom (in eV) of $Al_kTi_lNi_m$ ($k + l + m = 4$) clusters as a function of the atomic composition.

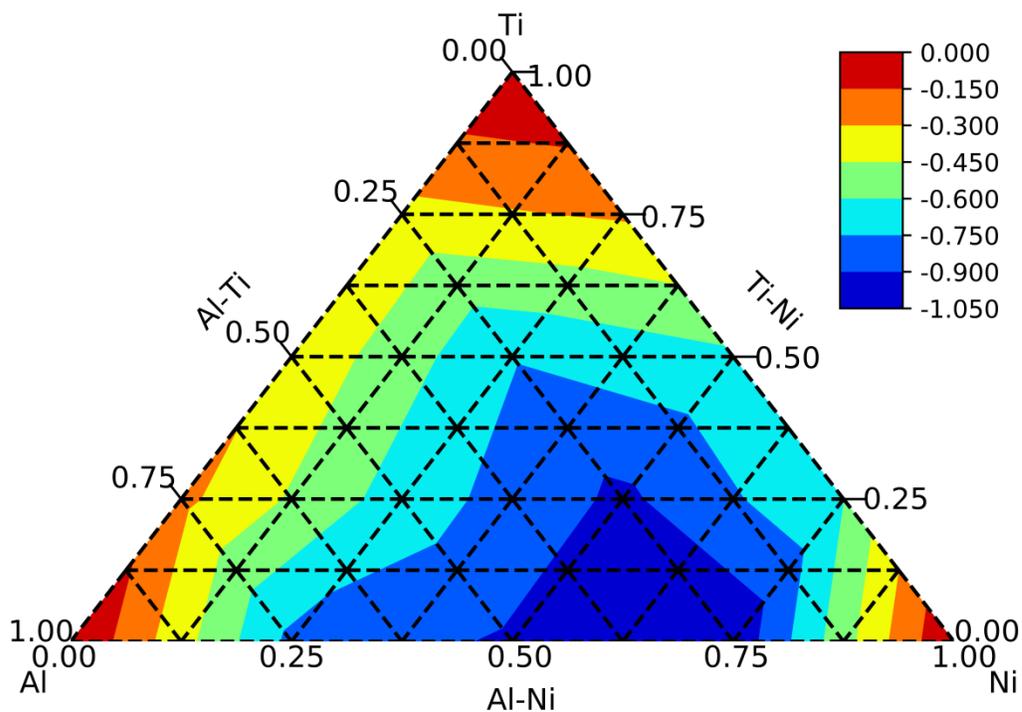

Fig. 4 Excess energy (in eV) of Al$_k$Ti$_l$Ni$_m$ ($k + l + m = 4$) clusters as a function of the atomic composition.

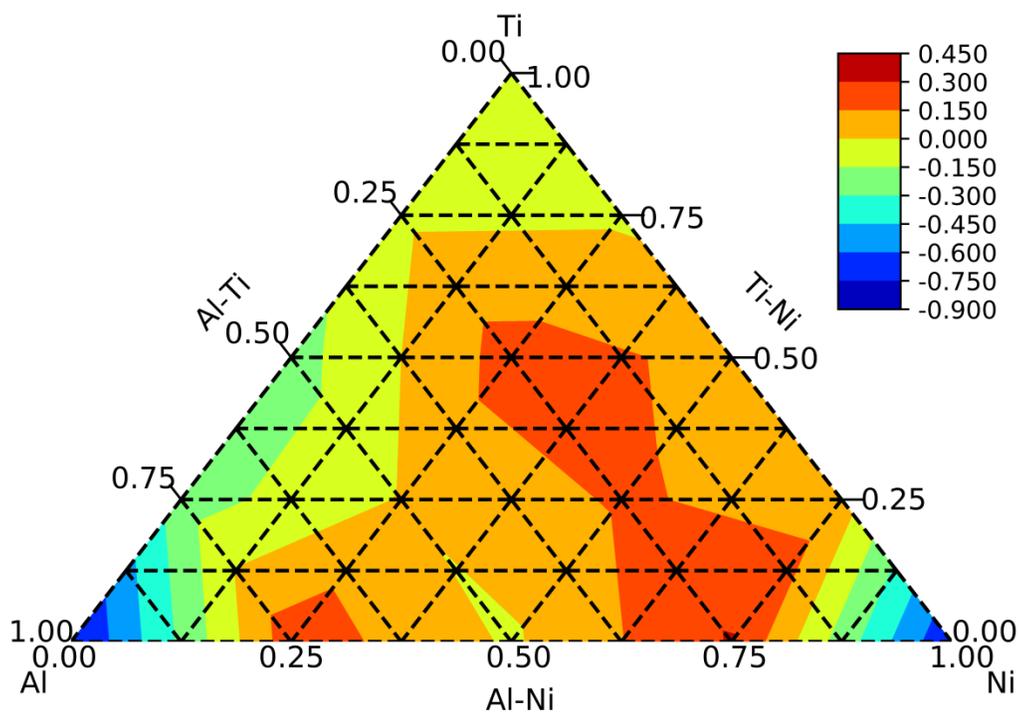

Fig. 5 Second order difference energy (in eV) of Al$_k$Ti$_l$Ni$_m$ ($k + l + m = 4$) clusters as a function of the atomic composition.

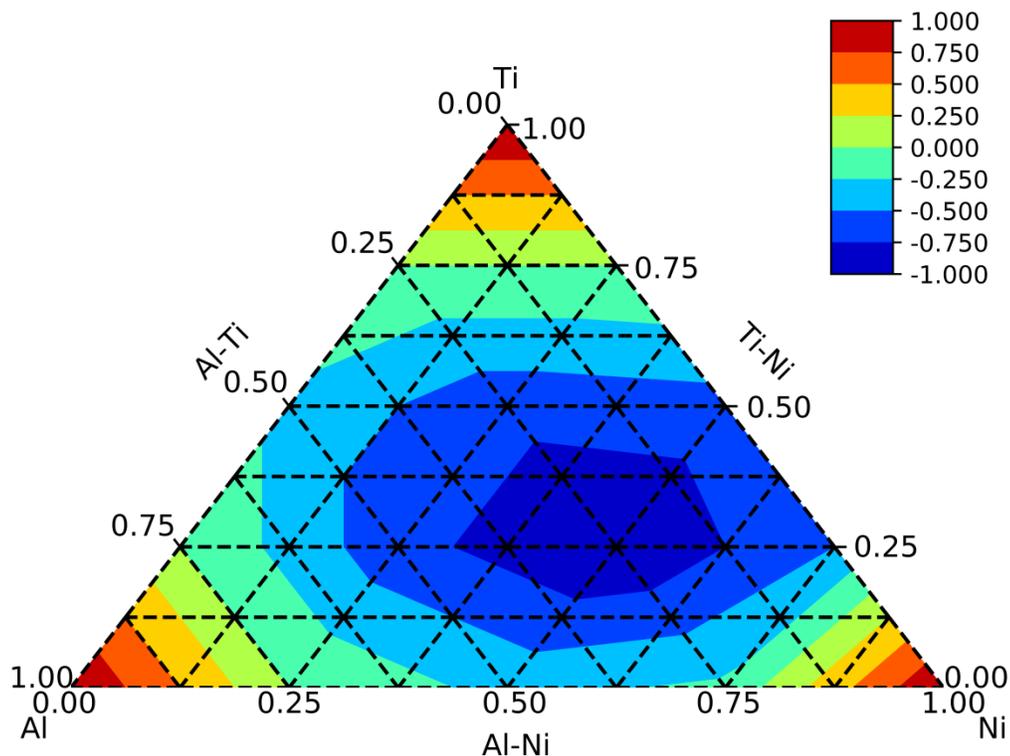

Fig. 6 Chemical order parameter (σ) of $Al_kTi_lNi_m$ ($k + l + m = 4$) clusters as a function of the atomic composition.

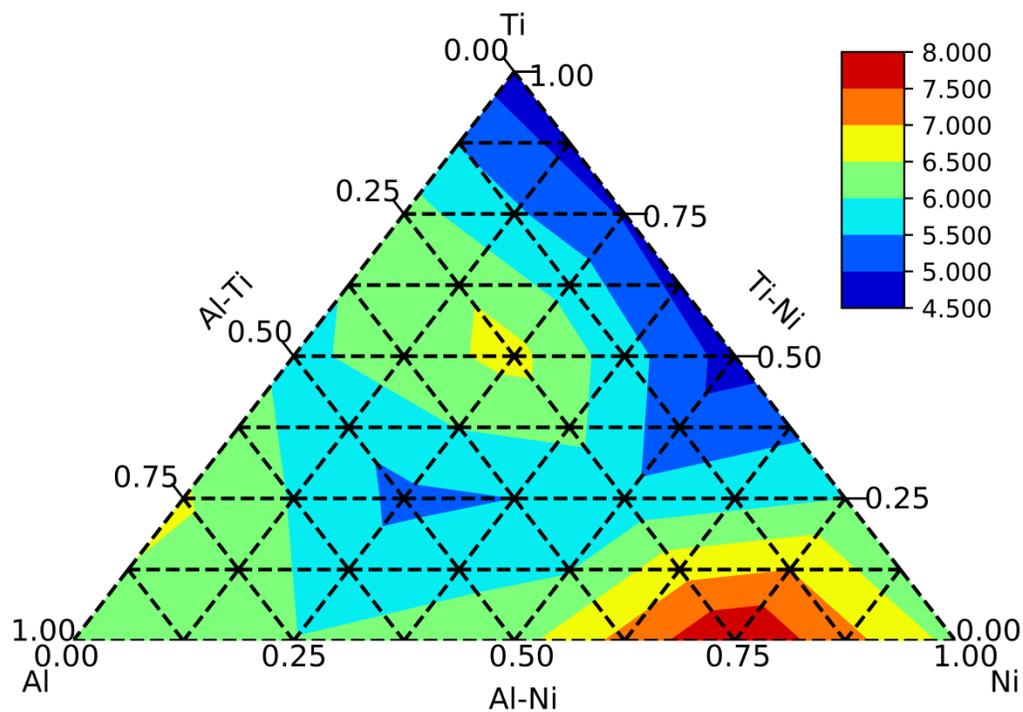

Fig. 7 Ionisation potential (IP) (in eV) of $Al_kTi_lNi_m$ ($k + l + m = 4$) clusters as a function of the atomic composition.

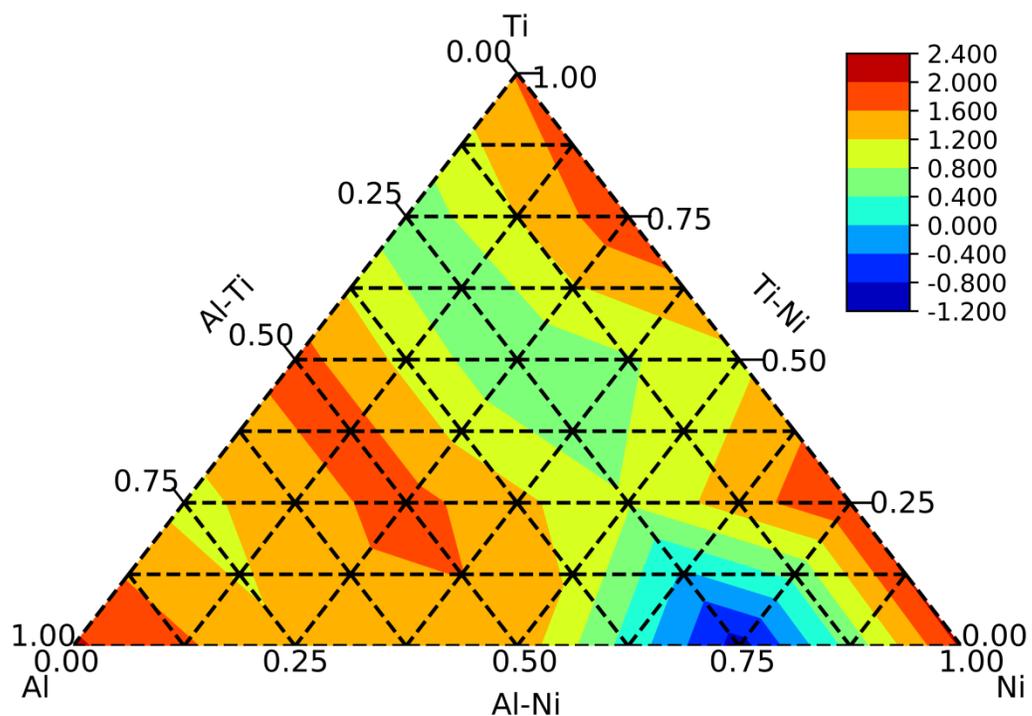

Fig. 8 Electron affinity (EA) (in eV) of $Al_kTi_lNi_m$ ($k + l + m = 4$) clusters as a function of the atomic composition.

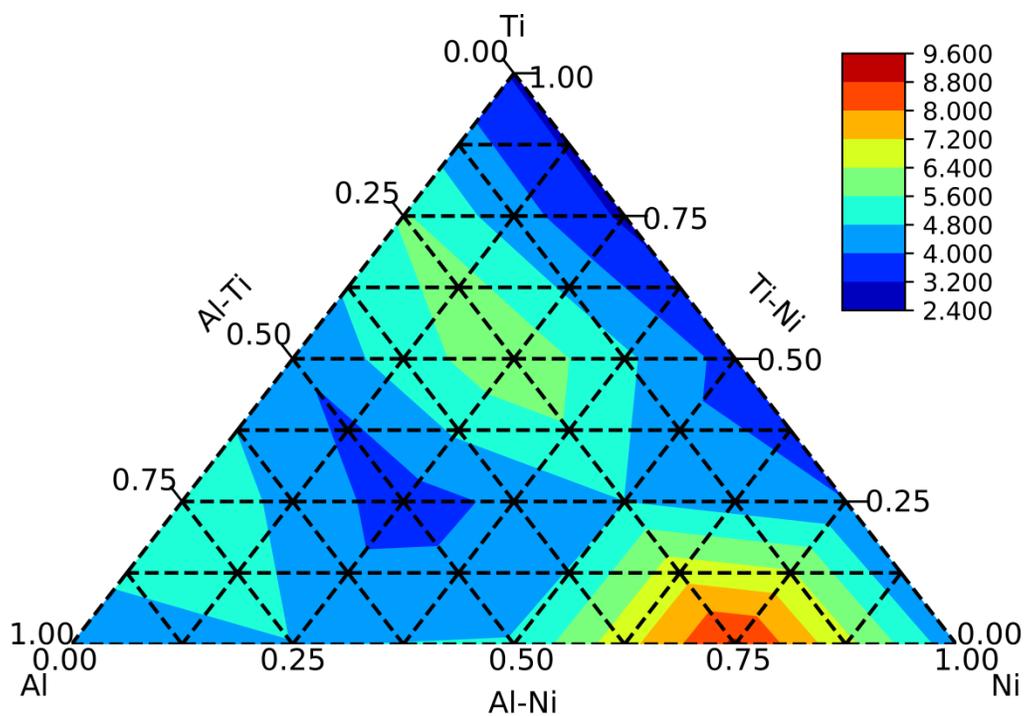

Fig. 9 Global hardness ($\eta$) (in eV) of $Al_kTi_lNi_m$ ($k + l + m = 4$) clusters as a function of the atomic composition.

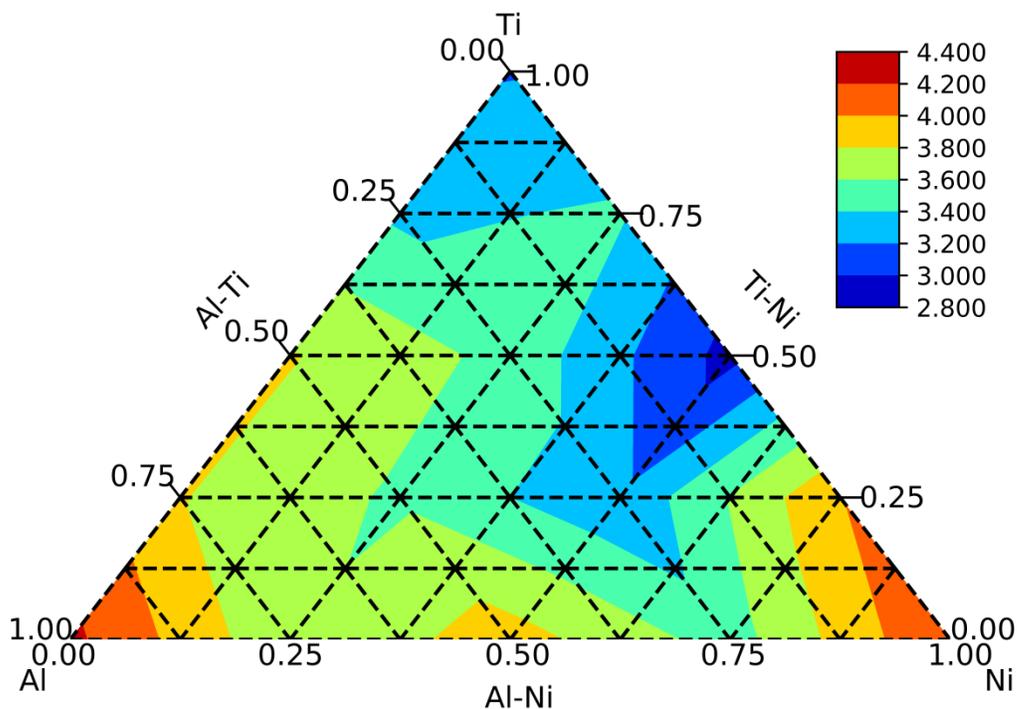

Fig. 10 Mulliken electronegativity ($\xi$) (in eV) of Al$_k$Ti$_l$Ni$_m$ ($k + l + m = 4$) clusters as a function of the atomic composition.

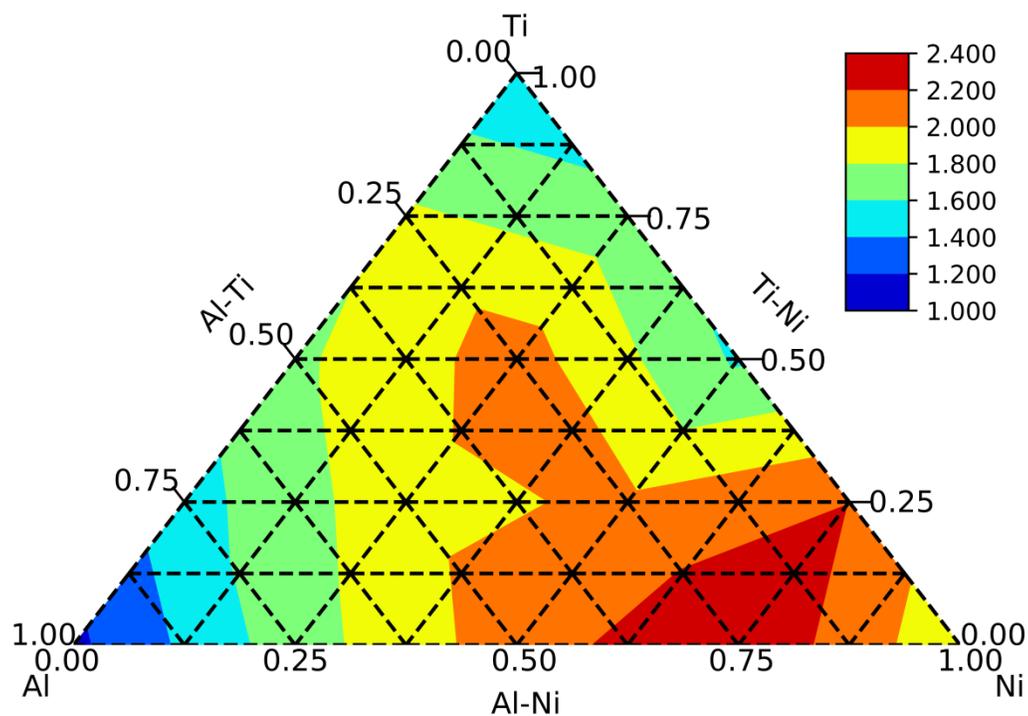

Fig. 11 HOMO-LUMO energy gaps (in eV) of Al$_k$Ti$_l$Ni$_m$ ($k + l + m = 4$) clusters as a function of the atomic composition.

## 4. Conclusion

We have calculated the stability, geometric and electronic properties of ternary alloy $Al_kTi_lNi_m$ ($k + l + m = 4$) clusters by using a two-stage basin-hopping genetic algorithms with density functional method at the SVWN/3–21G level of theory (for the first stage) and B3LYP/6–311G* level of theory (for the second stage). In $Al_kTi_lNi_m$ clusters, Al atoms incline to have the exposed positions which are lower coordinated while Ti atoms prefer to occupy the higher coordinated position or center position. Clusters with a high concentration of Al exhibit higher interatomic distance while those with a high concentration of Ni display smaller interatomic distance. Ni-rich clusters not only display a more minimum value of binding energy and excessive energy, it also exhibits a maximum value of second difference energy and HOMO-LUMO energy gap. Among all the clusters, $Al_1Ni_3$ cluster possesses a maximum value of ionization potential and global hardness. Analysis based on chemical order parameter indicates that ternary $Al_kTi_lNi_m$ clusters favor mixing rather than segregation. Although some general tendencies have been derived, further theoretical and experimental investigations of these Al-Ti-Ni clusters are required. It is hoped that the present work of small ternary $Al_kTi_lNi_m$ clusters able to act as a starting point for further research of physical, chemical and magnetic properties of ternary clusters, especially the magnetic properties of this particular clusters that not yet been investigated so far.

**Acknowledgments**

KPW acknowledges the financial support from the MyPhD scheme by the Malaysian Ministry of Higher Education (Kementerian Pendidikan Tinggi Malaysia). We acknowledge Prof. S. K. Lai from NCU (National Central University, Taiwan) for his tremendous support, crucial advice and provision of computational resources.